\documentstyle[sprocl]{article}

\input psfig.sty

\def\be{\begin{equation}}
\def\ee{\end{equation}}
\def\bea{\begin{eqnarray}}
\def\eea{\end{eqnarray}}
\def\beq{\begin{equation}}
\def\seq{\end{equation}}
\def\beqs{\begin{eqnarray}}
\def\seqs{\end{eqnarray}}

\def\D{\Delta_0}
\def\w{\omega}
\def\ve{\varepsilon}

\begin{document}

\title{ANISOTROPIC THERMODYNAMICS AND $T/\sqrt H$ SCALING
OF $d$-WAVE
SUPERCONDUCTORS IN THE VORTEX STATE.
}

\author{I. Vekhter$^1$,  P. J. Hirschfeld$^2$, J. P. Carbotte$^3$, and
E. J. Nicol$^1$}

\address{${}^1$Department of Physics, University of Guelph, Guelph,
        Ont. N1G 2W1, Canada\\
        ${}^2$Department of Physics, University of Florida,
        Gainesville, FL 32611, USA\\
        ${}^3$Department of Physics, McMaster University,
        Hamilton, Ont.  L8S 4M1, Canada
}


\maketitle\abstracts{  The density of states of a
2D $d$-wave superconductor in the vortex state
with applied magnetic field $\bf H$ in the plane
is shown to exhibit fourfold oscillations
as a function of the angle of the field with
respect to the crystal axes.  We find further
that the frequency dependence of the density of states and the
temperature dependence of transport coefficients obey different
power laws, thus  leading to different magnetic scaling functions,
for field in the nodal and anti-nodal direction.
We discuss the consequences of this anisotropy for measurements
of the specific heat.
}

Some of the first results
supporting the $d$-wave symmetry of
the energy gap in the hole-doped high-T$_c$ cuprates
came
 from
measurements of the specific heat in the vortex state \cite{moler}.
This has triggered a renewed interest in the
properties of
unconventional superconductors in the mixed state
\cite{phillips,junod,chiao}.

Volovik\cite{volovik} showed that in a magnetic field $H\ll H_{c2}$
the density of states (DOS) of a superconductor with  lines of nodes
is dominated
by the extended quasiparticle states, 
rather than by the bound states in the vortex
cores as for an $s$-wave gap. 
In a semiclassical treatment, the
energy of a quasiparticle with momentum {\bf k} at position {\bf r}
is shifted by
$\delta\omega_{\bf k}({\bf r})={\bf v}_s\cdot {\bf k}$,
where 
${\bf v}_s({\bf r})$
can be
approximated by  the velocity
field
around a single vortex.
Physical quantities then depend on {\bf r} and have
to be averaged over a unit cell of the vortex lattice.
This method has been successful in describing the
thermal and transport properties of the vortex state
with the field {\bf H} perpendicular to the CuO$_2$ layers 
\cite{volovik,kubert,kubert2,vekhter,barash}.

Recently we have argued that the same approach
remains qualitatively valid
in YBa$_2$Cu$_3$O$_{7-\delta}$
and other not too anisotropic materials
for {\bf H} parallel 
to the layers and have computed the DOS\cite{us}.
We 
assume an order parameter
$\Delta_{\bf k}=\Delta_0\cos2\phi$ over a
cylindrical Fermi surface parameterized by the angle $\phi$.
The field ${\bf H}$ is applied in the a-b plane, at an angle
$\alpha$ to the x-axis.
The flow field of a vortex, ${\bf v}_s$,
is elliptical due to the anisotropy
of the penetration depth, $\ve=\lambda_c/\lambda_{ab}$,
but after the rescaling of the c-axis
${\bf v}_s=\hbar\hat\beta/2mr$ is isotropic.\cite{us} Here
$r$ is the distance from the center of the vortex, and 
$\hat\beta$  is a unit vector along the current.
Approximating the unit cell of the vortex lattice by a circle of radius
$R=a^{-1}\sqrt{\Phi_0/\pi H}$, where $a\sim 1$ is a geometric constant,
we obtain
$\delta\omega_{\bf k}({\bf r})={E_H\sin\beta\sin(\phi-\alpha)/\rho}$,
where
$\vec\rho={\bf r}/R$ and a typical magnetic energy is defined as
$E_H={a} v^*\sqrt{\pi H/\Phi_0}/2$.
In London theory
$v^*=v_f/\sqrt{\ve}$, where $v_f$ is
the Fermi velocity in the plane.

For undoped $YBCO$,
$\ve\approx 5.3$ \cite{junod}, and the anisotropy
increases with deoxyge- nation\cite{basov}.
Taking
$v_f=1.2\times 10^7$cm/s \cite{chiao}
we estimate 
$E_H \approx 7.9a \sqrt H$ KT$^{-1/2}$ for the undoped samples, while 
for ${\bf H}\| c$ $E_H^c\equiv (v_F/v^*)E_H\approx 18.2 a\sqrt H$ KT$^{-1/2}$.
For in-plane field $\delta\omega_{\bf k}\propto\sin(\phi-\alpha)$ 
and the DOS $N(\omega,\alpha)$ depends
on the angle between the field and the direction
of
the gap nodes, giving\cite{volovik2,us} 
$N(0,\alpha)/N_0=2\sqrt{2} E_H{\rm max} (|\sin\alpha|,|\cos\alpha|)/(\pi
\Delta_0)$,
where $N_0$ is the normal state DOS.
Note that $N(0,\alpha)\propto\sqrt H$.
For $\w, E_H \ll \D$ the density of states
$N(\w,\alpha)\simeq (N_1(\w,\alpha)+N_2(\w,\alpha))/2$,
where
\beq
{N_i(\w, \alpha)\over N_0}=
\left\{
\begin{array}{ll} 
{\w\over\D}\left(1+{1\over 2x^2}\right), &
	\mbox{if } x=\w/E_i\ge 1;\\
	{E_i\over\pi\D x}\left[ (1+2x^2)\arcsin x+ 3x\sqrt{1-x^2}\right],
	& \mbox{if } x\le 1,
\end{array}\right.
\seq
for $i=1,2$, $E_1=E_H|\sin(\pi/4-\alpha)|$ and 
$E_2=E_H|\cos(\pi/4-\alpha)|$. \cite{us}
In particular
\beq
{N(\w,\alpha)\over N_0}\approx\left\{
\begin{array}{ll} {2\sqrt2 E_H\over \pi\D}
	\Bigl(1+{1\over 3} {\w^2\over E_H^2}\Bigr)
	& \mbox{for  } \alpha=0;\\
 {2 E_H\over \pi\D}+{\w\over 2\D} & \mbox{for  }\alpha=\pi/4
\end{array}\right..
\seq
The frequency dependence of $N(\w,\alpha)$
follows different power laws
for the field along a node or an anti-node, and  
consequently the specific heat coefficient $C/T$,
NMR relaxation time  $T_1T$ and other quantities exhibit
fourfold oscillations and a $T$ or $T^2$ behavior depending on the
direction of the field.

We now compute the
 low temperature specific heat $C(T,H)$
as in Ref.\cite{kubert}, and show the result
in Fig. 1a. Here $\gamma_n=\pi^2 N_0/3$,
and we have used $\ve =7$ and $E_H=0.1\D$. For $a=1$ and $\D =200$K  
this corresponds to
$H\simeq 6.5 T$. Taking\cite{moler} $\gamma_n =20$mJ/mol K$^2$, the
amplitude of the oscillations in $C/\gamma_n T$  for ${\bf H}\| ab$
at $T=0$ is
0.5 mJ/mol K$^2$, close to a previous estimate\cite{volovik2}. 
This amplitude is reduced as
$T$ increases: at $T=0.01\D\simeq2$K, it is 40\%
of the $T=0$ value. This can explain why the oscillations
have not been found in the one measurement done for two
orientations of the field\cite{moler}.
In an orthorhombic system the induced $s$-wave component of the gap shifts
the position of the DOS minimum away from the $\pi/4$ direction,
and in a heavily twinned crystal, such as used in Ref.\cite{moler},
this further suppresses
the amplitude of the oscillations.

For $T\ll E_H$,  
 $C/T$ varies as $T$ and $T^2$
for ${\bf H}\|node$ and ${\bf H}\|antinode$  respectively.
There exists a regime $E_H\ll T\ll E_H^c$ where
the anisotropy is washed out, $C({\bf H}\| ab)/T\propto T$ but
$C({\bf H}\| c)/T \simeq const$.
This observation can help resolve some of the disagreement
between the 
specific heat data obtained in Refs.\cite{moler,phillips,junod}.
The results of measurements both on single crystals\cite{moler}
with ${\bf H}\|c$, and on polycrystalline samples\cite{phillips}
are well described by $C/T\propto \sqrt H$. 
 Note
that due to large anisotropy, the supercurrents are
nearly in
the a-b plane for almost all orientations of the grains
with respect to {\bf H}  \cite{kogan},
so that both experiments effectively measure
$C({\bf H}\| c)$. Since the measured specific heat is a sum of the
DOS dependent and ``background''  contributions
the analysis is rather involved.
Instead, Revaz {\it et al.}\cite{junod} analyzed
the anisotropy $\delta C=C({\bf H}\| c)-C({\bf H}\| antinode)$,
interpreting it as a pure vortex quantity. They found 
$\delta C/T$
temperature dependent,
which can be understood since
it becomes $T-$dependent
for $E_H^c\gg T\ge E_H$.

We now define $C/(TE_H)\equiv N_0 F_C(X)/\D$, where
$X=T/E_H$ is the scaling variable\cite{simon}. In the limit
$X\gg 1$ we have 
$F_C(X)=9\zeta(3)X+\ln 2/2X$,
similar to the  result for ${\bf H}\| c$ \cite{kopnin}.
In the opposite limit
\beq
\label{smallX}
F_C(X)=\cases{2\pi/3 + 9\zeta(3)X/2, & $X\ll 1$, ${\bf H}\| node$,\cr
	      2 \sqrt 2\pi/3 + 14\sqrt 2 \pi^3 X^2/45,
			& $X\ll 1$, ${\bf H}\| antinode$.}
\seq
$F_C(X)$ is shown in Fig.1b
for $\ve=5.3$.
For ${\bf H}\| c$ the crossover from $C/TE_H\sim const$ 
occurs at $X_c\sim 0.5 $, which was estimated \cite{phillips}
to be at $T/\sqrt H=6.5$ KT$^{-1/2}$, 
yielding
$E_H^c\approx 30\sqrt H $KT$^{-1/2}$.
The crossover from small to large $X$
in  $\delta C(T,H)$,
occurs at $X_{ab}\sim 0.15 $, but 
the predicted $X\ll 1$ behavior was not found above\cite{junod}
$T/\sqrt H\approx 0.55$KT$^{-1/2}$, which implies
$E_H\leq 4\sqrt H$ KT$^{-1/2}$. Note that the interpolation 
used in Ref.\cite{junod} gives a linear correction in $X$, rather 
than quadratic as in
Eq.(\ref{smallX}), to the
behavior at $X\ll 1$,  which has led to an underestimate
of the crossover scale.
 Notice also that
the crossover in $\delta C$
extends over a decade in $X$.

The experimental values for $E_H^c$ and $E_H$ are
within a factor of 2 of our estimates, and
$E_H^c/E_H$ is then about 3 times the predicted ratio of $\sqrt{\ve}$. 
These are quantitative
rather than qualitative differences, 
given the roughness of the estimates and the experimental uncertainties.
Also the constant $a$ in general is not the same
for ${\bf H}\|ab$ and ${\bf H}\|c$. Note that for
Josephson coupled planes, the anisotropy in the
the vortex lattice constants is $H$-dependent and differs 
from $\ve$\cite{bulaevskii2}.
Finally, since the elastic moduli of the
vortex lattice differ for 
 ${\bf H}\|ab$ and ${\bf H}\|c$,
the lattice contributes to $\delta C$ \cite{reeves,fetter,bulaevskii}.

Qualitative agreement between our estimates and the experimental results
suggests that the angular
oscillations of the specific heat are accessible at $T\le 2$K 
and $H\sim 10-20$T. Such measurements
would be a simple bulk probe of gap symmetry allowing one
to map out the position of the gap nodes.

\begin{figure}[h]
\psfig{figure=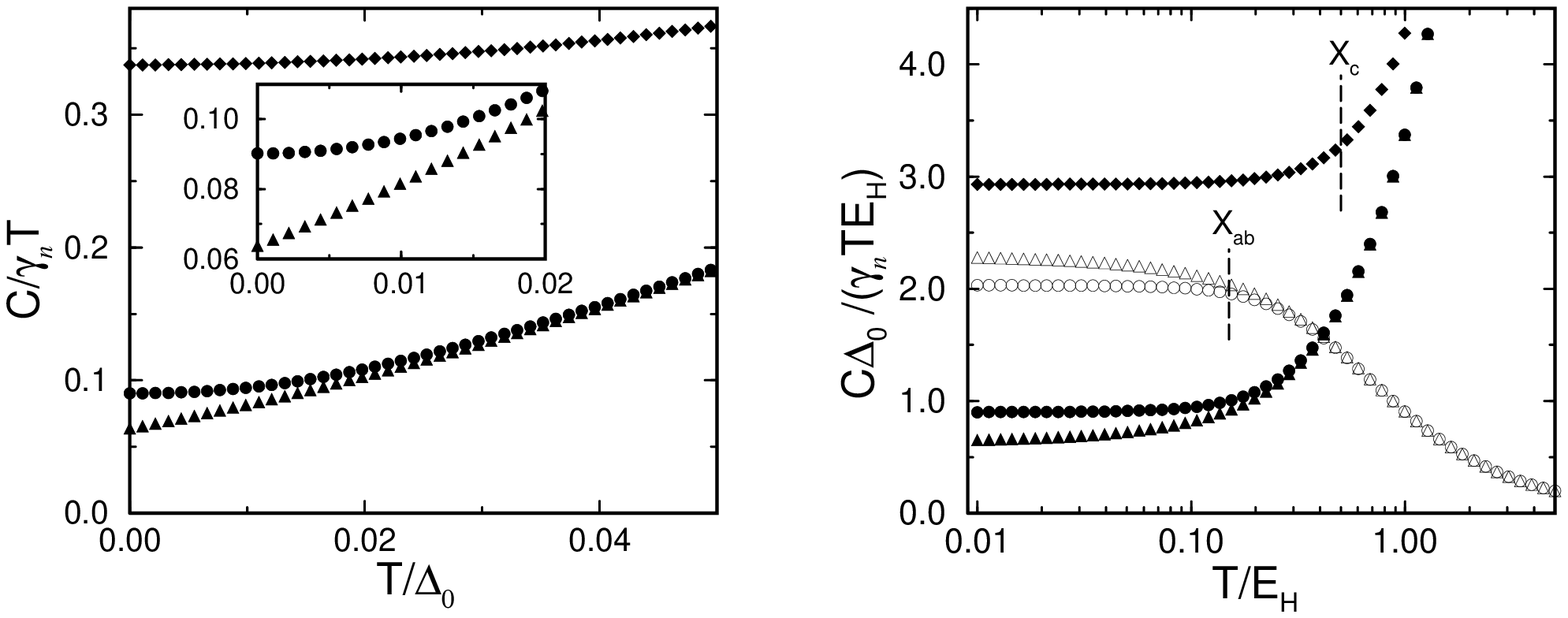,height=2.in}
\caption{a)Normalized specific heat for ${\bf H}\| c$ (diamonds),
${\bf H}\| antinode$ (circles), and ${\bf H}\| node$ (triangles), for
$\ve=7$ ($\delta\simeq 0.05$) and $E_H=0.1\D$; Inset: low-$T$ anisotropy;
b) Scaling function for $C({\bf H}\| c)$
(diamonds),
$C({\bf H}\| antinode)$ (full circles),
$C({\bf H}\| node)$ (full triangles),
$\delta C({\bf H}\| antinode)$ (open circles),
and $\delta C({\bf H}\| node)$ (open triangles).
}
\end{figure}

\section*{Acknowledgments}
Valuable communications with K.A. Moler, N. E. Phillips and G.E. Volovik are
acknowledged.
This research
has been supported in part by NSERC of Canada (EJN and JPC), CIAR (JPC)
and NSF/ AvH Foundation (PJH). EJN is a Cottrell Scholar of Research 
Corporation.


\section*{References}
\begin{thebibliography}{99}

\bibitem{moler} K. A. Moler {\it et al.}, Phys. Rev. Lett. {\bf 73}, 2744
(1994);
        Phys. Rev. {\bf B 55}, 3954 (1997); R. A. Fisher et al., 
Physica C {\bf 252}, 237 (1995).  

\bibitem{phillips} D. A. Wright {\it et al.}, to be published.

\bibitem{junod} 
	B. Revaz {\it et al.}, Phys. Rev. Lett. {\bf 80}, 3364 (1998).
\bibitem{chiao} M. Chiao {\it et al.}, 
        cond-mat/9810323.

\bibitem{volovik} G. E. Volovik, JETP Lett. {\bf 58}, 469 (1993).

\bibitem{kubert} C. K\"ubert and P. J. Hirschfeld, Sol. St. Comm. {\bf
105},
459 (1998).

\bibitem{kubert2} C. K\"ubert and P. J. Hirschfeld,
        Phys. Rev. Lett. {\bf 80}, 4963 (1998).

\bibitem{vekhter} I. Vekhter, J. P. Carbotte, and E. J. Nicol,
        cond-mat/9806033.

\bibitem{barash} Yu.S.Barash, V.P.Mineev, A.A.Svidzinskii,
        JETP Lett.{\bf 65}, 638 (1997).

\bibitem{us} I. Vekhter {\it et al.}, cond-mat/9809302.

\bibitem{basov} D. N. Basov {\it et al.} Phys. Rev. B {\bf 50}, 3511
(1994).

\bibitem{volovik2} G. E. Volovik, unpublished; in K. A. Moler {\it et
al.},
J. Phys. Chem. Solids {\bf 56}, 1899 (1995).

\bibitem{kogan} L.J.Campbell, M.M.Doria, V.G.Kogan,
Phys. Rev. B {\bf 38},
2439 (1988).

\bibitem{simon} S. H. Simon and P. A. Lee, Phys. Rev. Lett. {\bf 78},
	1548 (1997).

\bibitem{kopnin} N. B. Kopnin and G. E. Volovik, JETP Lett. {\bf 64},
690 (1996)


\bibitem{bulaevskii2} L. N. Bulaevskii and J. R. Clem, Phys. Rev. B {\bf
44},
                10234 (1991).
\bibitem{reeves} M. E. Reeves {\it et al.}, Phys. Rev. B {\bf 40}, 4573
(1989).

\bibitem{fetter} A. L. Fetter, Phys. Rev. B {\bf 50}, 13695 (1994). 
\bibitem{bulaevskii} L. N. Bulaevskii and M. P. Maley,
        Phys. Rev. Lett. {\bf 71}, 3541 (1993).



\end{thebibliography}
\end{document}